\documentclass[sigconf]{acmart}

\usepackage{balance}
\usepackage{booktabs} % For formal tables
\usepackage{pifont}
\usepackage{color}
\usepackage{adjustbox}

\usepackage{microtype} % must be loaded after all fonts
\usepackage{makecell}
\usepackage{graphicx}
\usepackage{subcaption}
\usepackage{amsmath}

\def\BibTeX{{\rm B\kern-.05em{\sc i\kern-.025em b}\kern-.08emT\kern-.1667em\lower.7ex\hbox{E}\kern-.125emX}}

\copyrightyear{2020}
\acmYear{2020}
\setcopyright{acmlicensed}
\acmConference[ASIA CCS '20]{Proceedings of the 15th ACM Asia Conference
on Computer and Communications Security}{October 5--9, 2020}{Taipei,
Taiwan}
\acmBooktitle{Proceedings of the 15th ACM Asia Conference on Computer and
Communications Security (ASIA CCS '20), October 5--9, 2020, Taipei, Taiwan}
\acmPrice{15.00}
\acmDOI{10.1145/3320269.3384728}
\acmISBN{978-1-4503-6750-9/20/10}

\newcommand{\sectionref}[1]{$\S$\ref{#1}}
\providecommand{\myparab}[1]{\smallskip\noindent\textbf{#1} }

\begin{document}

\title{Assessing the Privacy Benefits of Domain Name Encryption}

\author{Nguyen Phong Hoang}
\affiliation{%
  \institution{Stony Brook University}
}
\email{nghoang@cs.stonybrook.edu}

\author{Arian Akhavan Niaki}
\affiliation{%
  \institution{University of Massachusetts, Amherst}
}
\email{arian@cs.umass.edu}

\author{Nikita Borisov}
\affiliation{%
  \institution{University of Illinois at Urbana-Champaign}
}
\email{nikita@illinois.edu}

\author{Phillipa Gill}
\affiliation{%
  \institution{University of Massachusetts, Amherst}
}
\email{phillipa@cs.umass.edu}

\author{Michalis Polychronakis}
\affiliation{%
  \institution{Stony Brook University}
}
\email{mikepo@cs.stonybrook.edu}

\begin{abstract} 
{As Internet users have become more savvy about the potential for their
Internet communication to be observed, the use of network traffic encryption
technologies (e.g., HTTPS/TLS) is on the rise. However, even when encryption
is enabled, users leak information about the domains they visit via DNS
queries and via the Server Name Indication (SNI) extension of TLS. Two
recent proposals to ameliorate this issue are DNS over HTTPS/TLS (DoH/DoT)
and Encrypted SNI (ESNI).
In this paper we aim to assess the privacy benefits of
these proposals by considering the relationship between hostnames and IP
addresses, the latter of which are still exposed. We perform DNS queries from
nine vantage points around the globe to characterize this relationship. We
quantify the privacy gain offered by ESNI for different hosting and CDN providers
using two different metrics, the \emph{k}-anonymity degree due to co-hosting
and the dynamics of IP address changes. We find that 20\% of the domains
studied will not gain any privacy benefit since they have a one-to-one mapping
between their hostname and IP address. On the other hand, 30\% will gain a
significant privacy benefit with a \emph{k} value greater than 100, since
these domains are co-hosted with more than 100 other domains. Domains whose
visitors' privacy will meaningfully improve are far less popular, while for
popular domains the benefit is not significant. Analyzing the dynamics of IP addresses
of long-lived domains, we find that only 7.7\% of them change their hosting IP
addresses on a daily basis. We conclude by discussing potential approaches for
website owners and hosting/CDN providers for maximizing the privacy benefits
of ESNI.}
\end{abstract}

\begin{CCSXML}
<ccs2012>

  <concept>
    <concept_id>10003033.10003083.10011739</concept_id>
    <concept_desc>Networks~Network privacy and anonymity</concept_desc>
    <concept_significance>500</concept_significance>
  </concept>

  <concept>
    <concept_id>10003033.10003079.10011704</concept_id>
    <concept_desc>Networks~Network measurement</concept_desc>
    <concept_significance>500</concept_significance>
  </concept>

  <concept>
    <concept_id>10002978.10002991.10002995</concept_id>
    <concept_desc>Security and privacy~Privacy-preserving protocols</concept_desc>
    <concept_significance>500</concept_significance>
  </concept>

</ccs2012>
\end{CCSXML}

\ccsdesc[500]{Networks~Network privacy and anonymity}
\ccsdesc[500]{Networks~Network measurement}
%\ccsdesc[500]{Security and privacy~Privacy-preserving protocols}
\keywords{Domain name privacy; DNS over HTTPS (DoH); DNS over TLS (DoT);
Encrypted SNI (ESNI); active DNS measurement.}

\maketitle

\section{Introduction}
\label{sec:introduction}

As users become more aware of the importance of protecting their online
communication, the adoption of TLS is increasing~\cite{Kotzias:IMC18:TLS}. It
is indicative that almost 200M fully-qualified domain names (FQDNs) support
TLS~\cite{letsencrypt-report}, while Let's Encrypt~\cite{letsencrypt-paper}
has issued a billion certificates as of February 27,
2020~\cite{letsencrypt-billion}. Although TLS significantly improves the
confidentiality of Internet traffic, on its own it cannot fully protect user
privacy, especially when it comes to monitoring the websites a user visits.

Currently, visited domain names are exposed in both~i) DNS requests, which
remain unencrypted, and~ii) the Server Name Indication (SNI)
extension~\cite{rfc6066} during the TLS handshake. As a result, on-path
observers can fully monitor the domain names visited by web users through
simple eavesdropping of either DNS requests or TLS handshake traffic. Several
recent proposals aim to improve the security and privacy of these two
protocols. Specifically,
DNS over HTTPS (DoH)~\cite{rfc8484} and
DNS over TLS (DoT)~\cite{rfc7858} aim to
preserve the integrity and confidentiality of DNS resolutions against threats
``below the recursive,'' such as DNS poisoning~\cite{holdonDNS}, while
Encrypted Server Name Indication (ESNI)~\cite{rfc-draft-ietf-tls-esni-03} aims
to prevent ``nosy'' ISPs and other on-path entities from observing the actual
visited domain of a given TLS connection.

In this paper, we quantify the potential improvement to user privacy that a
full deployment of DoH/DoT and ESNI would achieve in practice, given that
destination IP addresses still remain visible to on-path observers. Although
it is straightforward to reveal a user's visited site if the destination IP
address hosts only that particular domain, when a given destination IP address
serves many domains, an adversary will have to ``guess'' which one is being
visited.

We use two properties to quantify the potential privacy benefit of ESNI,
assuming the provider of the DoH/DoT server used is fully trusted (as it can
still observe all visited domains): the \emph{k}-anonymity property and the
dynamics of hosting IP addresses. Assuming that \emph{k} different websites
are co-hosted on a given IP address (all using HTTPS with ESNI supported), the
privacy of a visitor to one of those sites increases as the number of
\emph{k-1} other co-hosted sites increases. In addition, the more dynamic the
hosting IP address is for a given site, the higher the privacy benefit of its
visitors is, as the mapping between domain and hosting IP address becomes less
stable, and thus less predictable.

To quantify these two properties, we conducted active DNS measurements to
obtain the IP addresses of an average of 7.5M FQDNs per day drawn from lists
of popular websites~\cite{alexa, majestic} (\sectionref{sec:method}). To
account for sites served from content delivery networks (CDNs) which may
direct users differently based on their location, we performed name resolutions
from nine locations around the world: Brazil, Germany, India, Japan, New
Zealand, Singapore, United Kingdom, United States, and South Africa. Our
measurements were conducted in two months to investigate how much a network
observer can learn about the domains visited by a user based solely on the IP
address information captured from encrypted traffic.

% While conducting this study, we faced the challenge of dealing with poisoned
% DNS responses caused by censorship leakage from China, as these poisoned
% responses would have a negative impact on our analyses if not excluded. Using
% the technique described in~\sectionref{sec:data_sanitization}, we detected
% more than 21K poisoned domains that otherwise would have been attributed as
% co-hosted with other legitimate domains. By comparing our results with other
% public datasets, we discovered that these poisoned domains have affected all
% these previous datasets.

We find that 20\% of the domains studied will not benefit at all from ESNI,
due to their stable one-to-one mappings between domain name and hosting IP
address. For the rest of the domains, only 30\% will gain a significant
privacy benefit with a \emph{k} value greater than 100, which means an
adversary can correctly guess these domains with a probability lower than 1\%.
The rest 50\% of the domains can still gain some privacy benefits, but at a
lower level (i.e., $2\leq k\leq 100$). While sophisticated website
fingerprinting attacks based on several characteristics of network packets
(e.g., timing and size~\cite{Wright2009NDSS, Lu2010ESORICS, Panchenko2011WPES,
fingerprinting16, HayesUSEC16, MiladCCS17}) can be used to predict the visited
domains, our study aims to provide a lower bound of what an attacker can
achieve.

Moreover, we observe that sites hosted by the top-ten hosting
providers with the highest privacy value ($k>500$) are far less popular
(\sectionref{sub:quantify_privacy}). These are often less well-known sites
hosted on small hosting providers that tend to co-locate many websites on a
single IP or server. In contrast, the vast majority of more popular sites
would gain a much lower level of privacy. These sites are often hosted by
major providers, including Cloudflare ($k=16$), Amazon ($3\leq k\leq 5$),
Google ($k=5$), GoDaddy ($k=4$), and Akamai ($k=3$).

In addition, we find that frequently changing IP addresses (at least once a
day) are limited to only 7.7\% of the domains that we were able to resolve
each day of our study. As expected, dominant providers in terms of more
dynamic IP addresses include major CDN providers, such as Amazon, Akamai, and
Cloudflare (\sectionref{sec:ip_stability}).

Finally, we validate and compare our main findings by repeating part of our
analysis using two different public DNS datasets
(\sectionref{sec:comparison}), and provide recommendations for both website
owners and hosting/CDN providers on how to maximize the privacy benefit
offered by the combination of DoH/DoT and ESNI (\sectionref{sec:discussion}).
In particular,
website owners may want to seek hosting services from---the unfortunately quite
few---providers that maximize the ratio between co-hosted domains per IP
address, and minimize the duration of domain-to-IP mappings. Hosting
providers, on the other hand, can hopefully aid in maximizing the privacy
benefits of ESNI by increasing the unpredictability of domain-to-IP mappings.

\section{Background}
\label{sec:background}

In this section, we provide an overview of the shortcomings of DNS and TLS
when it comes to user privacy, along with the suggested improvements of DNS
over HTTPS/TLS (DoH/DoT) and Encrypted Server Name Indication (ESNI).

\subsection{DNS and DoH/DoT}
\label{sec:dns}

%Communications over the Internet are made possible via the Internet
%Protocol~\cite{rfc791} in which each connected device is assigned with a
%unique identifier called IP address. However, IP addresses are formatted in a
%way that is difficult to remember (i.e., numeric for IPv4 or alphanumeric for
%IPv6). Domain Name System (DNS)~\cite{rfc1034} is introduced to eliminate the
%need for humans to memorize IP addresses. As a phonebook of the Internet, DNS
%plays an indispensable role in linking human-rememberable domain names to
%their associated machine-routable IP addresses. Almost every network
%communication on the Internet nowadays starts with a DNS lookup. For
%instance, when a user visits a website, her browser first needs to send a DNS
%lookup request to a DNS resolver. Once the browser receives a DNS response
%containing the corresponding IP address of the website, it can then initiate
%a connection to fetch the site's content.

The DNS protocol exposes all requests and responses in plaintext, allowing
anyone with the privilege to monitor or modify a user's network traffic to
eavesdrop or tamper with it. For example, a man-on-the-side attacker can send
spoofed DNS responses to redirect a victim to malicious
websites~\cite{holdonDNS}, while state-level organizations can manipulate DNS
responses to disrupt connections for censorship purposes. The DNS Security
Extensions (DNSSEC)~\cite{rfc2065} were introduced in 1997 to cope with these
and other security issues by assuring the integrity and authenticity of DNS
responses (but not their confidentiality). However, DNSSEC is still not widely
deployed due to deployment difficulties and compatibility
issues~\cite{Dai:CANS16:DNSSEC, Chung:Usenix17:DNSSEC, Hao:Usenix18:CDNSEC}.

As an attempt to enhance the security and privacy of the DNS protocol, two
emerging DNS standards were recently proposed: DoH~\cite{rfc8484} and
DoT~\cite{rfc7858}. These technologies aim to not only ensure the integrity
and authenticity of DNS traffic, but also its confidentiality to some extent.
Using DoH/DoT, all DNS queries and responses are transmitted over TLS,
ensuring their integrity against last-mile adversaries who would otherwise be
in a position to launch man-in-the-middle (MiTM) and man-on-the-side (MoTS)
attacks. In this work, we specifically characterize the protection of user
privacy from nosy ISPs and other last-mile entities provided by DoH/DoT.

Although the benefits of DoH/DoT against last-mile adversaries are clear, this
comes with the cost of \emph{fully trusting} a third-party operator of the
DoH/DoT resolver on which users have outsourced all their DNS
resolutions~\cite{Hoang2020:MADWeb}. Several companies already offer public
DoH/DoT resolvers, including Google~\cite{googleDoH, googleDoT} and
Cloudflare~\cite{cloudflare1111}. In fact, we later show
in~\sectionref{sub:quantify_privacy} that these two companies are also the
most dominant \emph{hosting} providers of domains in the top lists of popular
sites. Popular browsers have also started introducing support for DoH/DoT,
e.g., Mozilla Firefox since version 62~\cite{firefoxDoH} (which is now enabled
by default).

\subsection{SSL/TLS and ESNI}
\label{sec:tls}

%Secure Sockets Layer (SSL)~\cite{rfc6101} is a protocol that guarantees the
%confidentiality, integrity, and authenticity of communications over the
%insecure Internet environment. SSL was updated and evolved into Transport
%Layer Security (TLS) in 1999~\cite{rfc2246}. TLS consists of sophisticated
%cryptographic algorithms to make sure that traffic between two sides of a
%communication can be securely transmitted by safeguarding the information
%against MiTM eavesdroppers. Since the design of TLS and its working mechanism
%is too complicated and beyond the scope of this paper, we only discuss aspects
%of TLS that impact the privacy of web users.

During the TLS handshake~\cite{rfc2246}, the two communicating parties
exchange messages to acknowledge and verify the other side using digital
certificates, and agree on various parameters that will be used to create an
encrypted channel. In a client-server model, the client trusts a digital
certificate presented by the server as long as it has been signed by a trusted
certificate authority.

Ideally, private or sensitive information should be transmitted only after the
TLS handshake has completed. This goal can be easily achieved when a server
hosts only a single domain (known as IP-based virtual hosting). Name-based
virtual hosting, however, which is an increasingly used approach for enabling
multiple domains to be hosted on a \emph{single} server, necessitates a
mechanism for the server to know which domain name a user intends to visit
\emph{before} the TLS handshake completes, in order to present the right
certificate. The Server Name Indication (SNI) extension was introduced in 2003
as a solution to this problem. The SNI extension contains a field with the
domain name the client wants to visit, so that the server can then present the
appropriate certificate. Unfortunately, since this step is conducted prior to
the completion of the TLS handshake, the domain name specified in SNI is
exposed in plaintext. Consequently, all the privacy risks associated with the
traditional design of DNS discussed above also apply to the SNI extension. For
instance, Internet authorities in several countries have been relying on the
SNI field for censorship purposes~\cite{hoang:2019:measuringI2P,
zchai2019foci}.

ESNI has recently been proposed as part of TLS version
1.3~\cite{rfc-draft-ietf-tls-esni-03} to resolve the issue of SNI revealing
the domain visited by a user. Using ESNI, clients encrypt the SNI field
towards a given server by first obtaining a server-specific public key through
a well-known ESNI DNS record. Obviously, due to this reliance on DNS, any
privacy benefits of ESNI can only be realized when used in conjunction with
DoH/DoT---otherwise any last-mile observer would still be able to observe a
user's plaintext DNS queries and infer the visited TLS server. In September
2018, Cloudflare was among the first providers to announce support for ESNI
across its network~\cite{cloudflareESNI}.

\section{Threat Model}
\label{sec:threat_model}

We assume an idealistic future scenario in which both DoH/DoT and ESNI are
\emph{fully} deployed on the Internet. Under this assumption, an on-path
observer will only be able to rely on the remaining visible information, i.e.,
destination IP addresses, to infer the sites being visited by the monitored
users. The extent to which this inference can be easily made depends on i)
whether other domains are hosted on the same IP address, and ii) the stability
of the mapping between a given domain and its IP address(es).

The probability with which an adversary can successfully infer the visited
domain can be modeled using the \emph{k}-anonymity property, with \emph{k}
corresponding to the number of domains co-hosted on the same IP address. The
probability of a successful guess is inversely proportional to the value of
\emph{k}, i.e., the larger the \emph{k}, the more difficult it is for the
adversary to make a correct guess, thus providing increased user privacy.

The above threat model is oblivious to distinguishable characteristics among a
group of co-hosted websites, such as popularity ranking, site sensitivity, and
network traffic patterns. We should thus stress that the situation in practice
will be \emph{much more favorable} for the adversary. Even for a server with a
high \emph{k}, it is likely that not all \emph{k} sites will be equally
popular or sensitive. Although the popularity and sensitivity can vary from
site to site, depending on who, when, and from where is visiting the
site~\cite{Hoang2016:TACT}, an adversary can still consider the popularity and
sensitivity of the particular \emph{k} sites hosted on a given IP address to
make a more educated guess about the actual visited site.

Utilizing the ranking information of all domains studied, we model such an
adversarial scenario in~\sectionref{sub:quantify_weighted_privacy} and show
that our threat model based on \emph{k}-anonymity is still valid. In addition,
page-specific properties such as the number of connections towards different
third-party servers and the number of transferred bytes per connection can be
used to derive robust web page \emph{fingerprints}~\cite{encrhttp06,
Herrmann2009CCSW, touching12, peekaboo12, Cai2014CCS, fingerprinting14,
Patil:2019, Cui:AsiaCCS19}, which can improve the accuracy of attribution even
further. Although identifying a visited website among all possible websites on
the Internet by relying solely on fingerprinting is quite challenging,
applying the same fingerprinting approach for attributing a given connection
(and subsequent associated connections) to one among a set of \emph{k}
\emph{well-known} websites is a vastly easier problem.

Consequently, an on-path observer could improve the probability of correctly
inferring the actual visited website by considering the popularity and sensitivity
of the co-hosted domains on the visited IP address, perhaps combined with
a form of traffic fingerprinting. Although such a more powerful attack is outside the
scope of this work, as we show in the rest of the paper, our results already
provide a worrisome insight on how effective an even much less sophisticated
attribution strategy would be, given the current state of domain co-hosting.

\section{Methodology}
\label{sec:method}

In this section we review existing DNS measurement techniques and highlight
the data collection goals of our study. We then describe how we select domains
and vantage points to achieve these goals.

% as well as how we sanitize our data in the presence of censorship leakage from
% the Great Firewall of China. 

\subsection{Existing DNS Measurements}
\label{sec:dns_measurement}

%There are two main types of measurements for the collection of DNS-related
%data: passive and active measurements. In a \emph{passive measurement}, DNS
%data is obtained by an entity who has the privilege to capture DNS traffic
%from the network infrastructure under its control (e.g., networks of academic
%institutes or small organizations)~\cite{weimer2005passive}. 

Previous studies use passive measurement to observe DNS traffic on their
networks~\cite{weimer2005passive, Shue:2007, Dell'Amico:ACSAC17}. However,
passive data collection can suffer from bias depending on the time, location,
and demographics of users within the observed network. Passive data collection
can also raise ethical concerns, as data collected over a long period of time
can gradually reveal online habits of monitored users.

%In contrast, \emph{active measurement} technique involves purposely sending
%and receiving DNS queries and responses. Researchers can choose which domains
%to resolve depending on the goals of their study, thus having more control
%over the collected data. Although this approach sidesteps the privacy issues
%of passive DNS collection, it requires an increased amount of resources for
%operating a dedicated measurement infrastructure if the amount of domains
%need to be resolved is large~\cite{Rijswijk-Deij:2015}.
 
There are also prior works (by both academia and industry) that conducted
large-scale active DNS measurements for several purposes and made their
datasets available to the community~\cite{Kountouras2016, rapid7}. However,
these datasets have two common issues that make them unsuitable to be used
directly in our study. First, all DNS queries are resolved from a single
location (country), while we aim to observe localized IPs delivered by CDNs to
users in different regions. Second, although these datasets have been used in
many other studies, none of the prior measurements are designed to filter out
poisoned DNS responses (e.g., as a result of censorship leakage), which can
significantly affect the accuracy of the results and negatively impact data
analysis if not excluded. We discuss steps taken to sanitize these datasets in
Appendix~\ref{sec:data_sanitization}.

\subsection{Our Measurement Goals}
\label{sec:dataset}

Ideally, we would like to derive the mapping between all live domain names and
their IP addresses. Unfortunately, this is extremely challenging to achieve in
practice because there are more than 351.8 million second-level domain names
registered across all top-level domains (TLDs) at the time we compose this
paper~\cite{verisign.domains}, making it unrealistic to actively resolve all
of them with adequate frequency. Furthermore, not all domains host web
content, while many of them correspond to spam,
phishing~\cite{Pariwono:ASIACCS18, Quinkert:CNS19}, malware command and
control~\cite{Alowaisheq:NDSS19}, or parking pages registered during the
domain dropcatching process~\cite{Lauinger:usenix17}, which most users do not
normally visit. 

As we aim to study the privacy benefits of ESNI, we thus choose to focus on
active sites that are legitimately visited by the majority of web users. To
derive such a manageable subset of sites, we relied on lists of website
rankings, but did not consider only the most popular ones, as this would bias
our results. Instead, we expanded our selection to include as many sites as
possible, so that we can keep our measurements manageable, but at the same
time observe a representative subset of \emph{legitimately visited} domains on
the Internet.

\subsection{Domains Tested}

There are four top lists that are widely used by the research community:
Alexa~\cite{alexa}, Majestic~\cite{majestic}, Umbrella~\cite{cisco_umbrella},
and Quantcast~\cite{quantcast}. However, it is challenging to determine which
top list should be chosen, as recent works have shown that each top list has
its own issues that may significantly affect analysis results if used without
some careful considerations~\cite{Scheitle:2018:TopList, LePochat2019,
Rweyemamu2019}. For instance, Alexa is highly fluctuating, with more than 50\%
of domain names in the list changing every day, while Majestic is more stable
but cannot quickly capture sites that suddenly become popular for only a short
period of time. Pochat et al.~\cite{LePochat2019} suggest that researchers
should combine these four lists to generate a reliable ranking.

For this study, we generated our own list by aggregating domains ranked by
Alexa and Majestic from the most recent 30 days for several reasons. First,
these two lists use ranking techniques that are more difficult and costly to
manipulate~\cite{LePochat2019}. Second, they have the highest number of
domains in common among the four. We exclude domains from Quantcast because it
would make our observations biased towards popular sites only in the
US~\cite{LePochat2019}. Lastly, we do not use domains from Umbrella because
the list is vulnerable to DNS-based manipulation and also contains many
domains that do not host web content~\cite{LePochat2019, Rweyemamu2019}. To
this end, we studied a total of 13.6M domains with its breakdown shown in
Appendix~\ref{sec:tld_breakdown}.

\myparab{Data scope.} Although this subset of domains corresponds to about 4\%
of all domains in the TLD zone files, we argue that it is still adequate for
the goal of our study, i.e., determining whether the current state of website
co-location will allow ESNI to provide a meaningful privacy benefit.
Considering only this subset of domains may lead to an under-approximation of
the actual \emph{k}-anonymity offered by a given IP or set of IPs, as some
co-hosted domains may not be considered. This means that our results can be
viewed as a lower bound of the actual \emph{k}-anonymity degree for a given
visited IP address, which is still a desirable outcome.

As discussed in~\sectionref{sec:threat_model}, the popularity of a website,
along with other qualitative characteristics, can be used by an adversary to
improve attribution. Indeed, given that the long tail of domains that are left
out from our dataset mostly correspond to vastly less popular and even
unwanted or dormant domains~\cite{Szurdi:usenixsecurity14}, any increase in
\emph{k} they may contribute would in practice be rather insignificant, as
(from an attribution perspective) it is unlikely they will be the ones that
most web users would actually visit.

\subsection{Measurement Location and Duration}

Due to load balancing and content delivery networks, deriving \emph{all}
possible IP addresses for a given popular domain is very challenging. To
approximate this domain-to-IP mapping, we performed our own active DNS
measurements from several vantage points acquired from providers of Virtual
Private Servers (VPS). When choosing measurement locations, we tried to
distribute our vantage points so that their geographical distances are
maximized from each other. This design decision allows us to capture as many
localized IP addresses of CDN-hosted sites as possible. To that end, we run
our measurements from nine countries, including Brazil, Germany, India, Japan,
New Zealand, Singapore, United Kingdom, United States, and South Africa. Our
vantage points span the six most populous continents. From all measurement
locations mentioned above, we send DNS queries for approximately 7.5M domains
on a daily basis. When issuing DNS queries, we enabled the iterative flag in
the queries, bypassing local recursive resolvers to make sure that DNS
responses are returned by actual authoritative name servers. The results
presented in this work are based on data collected for a period of two months,
from February 24th to April 25th, 2019.

\section{Data Analysis}
\label{sec:data_analyses}

In this section we use two metrics, \emph{k}-anonymity and the
dynamics of hosting IP addresses, to quantify the privacy benefits offered by
different hosting and CDN providers. To verify the validity of our
\emph{k}-anonymity model, we also apply Zipf's law on the popularity ranking
of domains to account for a more realistic (i.e., more powerful) adversary.

\subsection{Single-hosted vs. Multi-hosted Domains}
\label{sec:cohosted_analysis}

\begin{figure}[t]
\centering
\includegraphics[width=0.95\columnwidth]{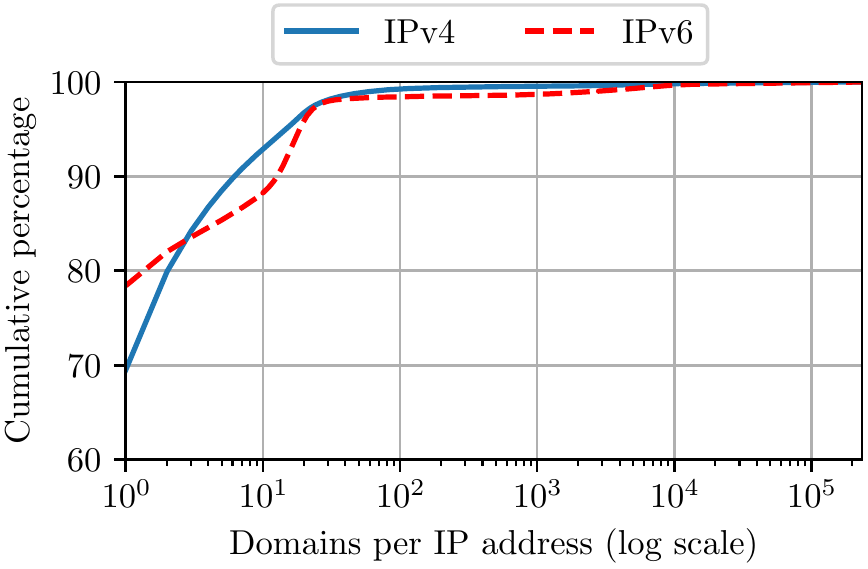}
\caption{Cumulative distribution function (CDF) of the number of domains
hosted per IP address, as a percentage of all observed IP addresses. About
70\% of all observed IPv4 addresses host only a single domain.}
\label{fig:ip2domains}
\end{figure}

Over a period of two months, from February 24th to April 25th, 2019, we
observed an average of 2.2M and 500K unique IPv4 and IPv6 addresses,
respectively, from our daily measurements. Of these IP addresses, 70\% of IPv4
and 79\% of IPv6 addresses host only a single domain, as shown in
Figure~\ref{fig:ip2domains}. This means that visitors of the websites hosted
on those addresses will not gain any meaningful privacy benefit with ESNI, due
to the one-to-one mapping between the domain name and the IP address on which
it is hosted. About 95\% of both IPv4 and IPv6 addresses host less than 15
domains.

When calculating the percentage of IPv6-supported sites, we find that less
than 15\% support IPv6. Regardless of the increasing
trend~\cite{Czyz:SIGCOMM14}, the future adoption of IPv6 is still
unclear~\cite{Colitti:PAM10}. Since the majority of web traffic is still being
carried through IPv4, in the rest of the paper we focus only on IPv4
addresses.

%While Figure~\ref{fig:ip2domains} shows that 70\% of destination IPs have a
%one-to-one mapping to domains, this does not necessarily mean that 70\% of
%\emph{domains} have a one-to-one mapping. We investigate this further in
%Section~\ref{sub:quantify_privacy}.

%Given the results of Figure~\ref{fig:ip2domains}, one may think that 70\% of
%the web will not gain any privacy benefits from a full deployment of ESNI.
%This is not the case, however, because the large growth in the use of CDNs and
%domain co-hosting---which motivated the introduction of ESNI in the first
%place---makes the accurate mapping of domain-to-IP address across different
%network locations more challenging. For instance, many popular domains with a
%large number of visitors often serve their content from many IP addresses via
%CDNs. These IP addresses may also serve any number of other domains, in which
%case ESNI can actually improve a user's privacy.

\begin{figure*}[t]
\centering
\includegraphics[width=0.854\textwidth]{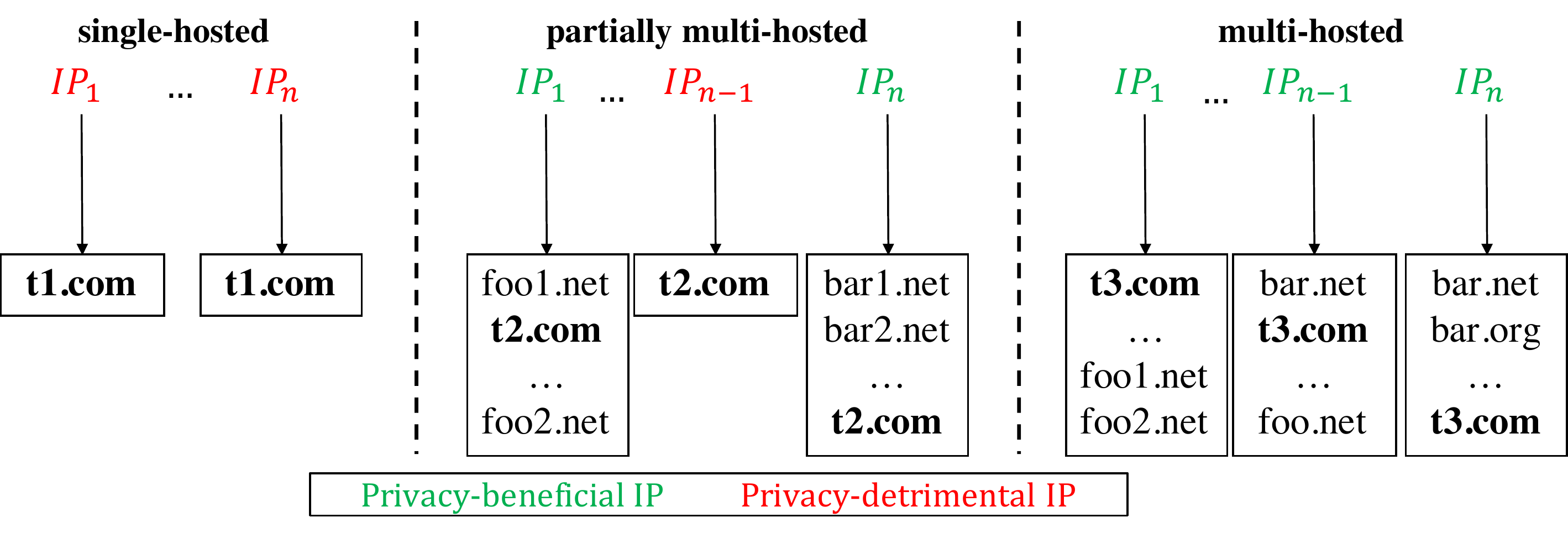}
\caption{Different types of domain hosting according to whether they can
  benefit from ESNI. Single-hosted domains are exclusively hosted on one or
  more IP addresses, and thus cannot benefit from ESNI. In contrast,
  multi-hosted domains are always co-hosted with more domains on a given IP
  address, and thus can benefit by ESNI.}
\label{fig:colocated_types}
\end{figure*}

Based on our measurements, we identify three main ways in which a domain may
be hosted, in terms of the IP addresses used and the potential privacy benefit
due to ESNI, as illustrated in Figure~\ref{fig:colocated_types}. In the
simplest case, a \emph{single-hosted} domain may be \emph{exclusively hosted}
on one or more IP addresses that do not serve any other domain, to which we
refer as \emph{privacy-detrimental} IP addresses
(Fig.~\ref{fig:colocated_types}, left). As there is no sharing of the IP
address(es) with other domains, an adversary can trivially learn which site is
visited based solely on the destination IP address. On the other hand, a
\emph{multi-hosted} domain (Fig.~\ref{fig:colocated_types}, right) may be
\emph{co-hosted} on one or more IP addresses that always serve at least one or
more other domains, to which we refer as \emph{privacy-beneficial} IP
addresses. Since the destination IP address always hosts multiple domains, an
adversary can only make a (possibly educated) guess about the actual domain a
given user visits, and thus multi-hosted domains always benefit to some extent
from ESNI---the more co-hosted domains on a given IP address, the higher the
privacy gain offered by ESNI.

Finally, there is a chance that a domain is hosted on a mix of
privacy-detrimental and privacy-beneficial IP addresses, which we call
\emph{partially multi-hosted} domains (Fig.~\ref{fig:colocated_types},
middle). In that case, only visitors to the subset of IP addresses that
co-host other domains will benefit from ESNI. Based on our measurements,
partially multi-hosted domains correspond to only a 0.3\% fraction (20K) of
all domains (daily average). Single-hosted domains comprise 18.7\% (1.4M) and
multi-hosted domains comprise 81\% (6M) of all domains.

The privacy degree of a partially multi-hosted domain depends on the
probability that a visitor gets routed to a privacy-beneficial IP of that
domain. In other words, a partially multi-hosted domain will mostly behave as
a multi-hosted domain if the majority of its IP addresses are
privacy-beneficial. In fact, we find that this is the case for more than
92.5\% of the partially multi-hosted domains studied. Based on this fact, and
given its extremely small number compared to the other two types, in the rest
of our paper we merge partially multi-hosted domains with the actual
multi-hosted domains, to simplify the presentation of our results.

Going back to Figure~\ref{fig:ip2domains}, based on the above breakdown, we
observe that 70\% of all IP addresses that host a single domain correspond
to 18.7\% of all domains, i.e., the single-hosted ones. On the other hand, the
81\% of multi-hosted domains are co-hosted on just 30\% of the IP addresses
observed.

\begin{figure}[t]
\centering
\includegraphics[width=0.95\columnwidth]{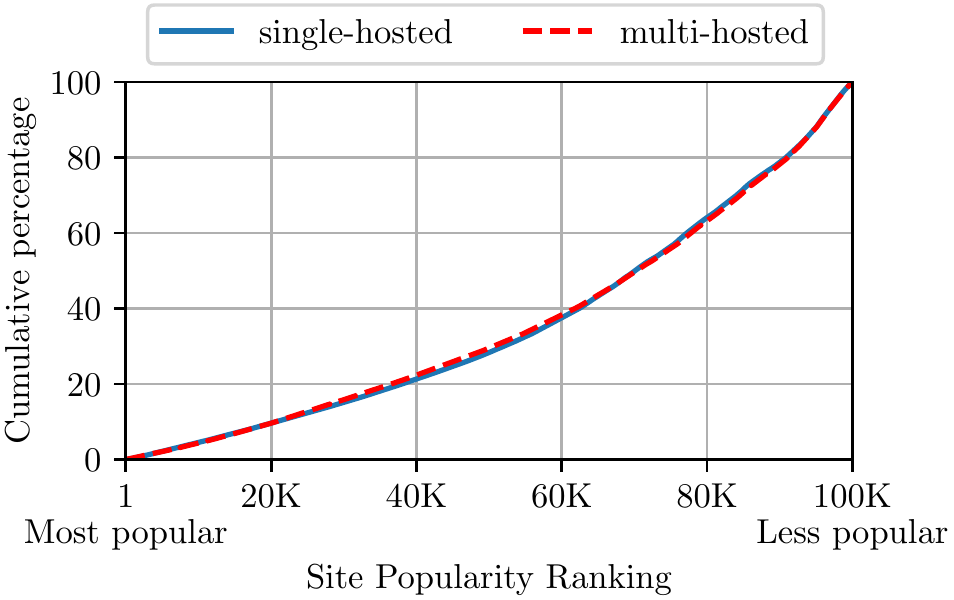}
\caption{CDF of the popularity ranking for single-hosted and multi-hosted
domains.}
\label{fig:popularity_distribution}
\end{figure}

Next, we analyze the popularity distribution of single-hosted and multi-hosted
domains to identify any difference in the user population of these two types
of domains.
Note that we only base our analysis on the ranking information provided by the
top lists to comparatively estimate the scale of the user base, and not for
absolute ranking purposes. More specifically, we only use the top 100K domains for
the analysis in Figure~\ref{fig:popularity_distribution}, since rankings lower
than 100K are not statistically significant, as confirmed by both top list
providers and recent studies~\cite{alexa.qa2, Rweyemamu2019}.
Figure~\ref{fig:popularity_distribution} shows that single-hosted and
multi-hosted domains exhibit a nearly identical distribution of popularity
rankings.

\subsection{Estimating the Privacy Benefit of Multi-hosted Domains}
\label{sub:quantify_privacy}

\begin{table}[t]
\centering
\caption{Top hosting providers offering the highest median \emph{k}-anonymity per IP
address.}
\begin{tabular}{rlrr}
\toprule
\small{Median} & \small{Organization}   & \small{Unique} & \small{Highest}\\
\small{\emph{k}} & \small{}   & \small{IPs} & \small{Rank}\\
\midrule
3,311 & AS19574 Corporation Service & 2 & 1,471   \\
2,740 & AS15095 Dealer Dot Com      & 1 & 80,965  \\
2,690 & AS40443 CDK Global          & 1 & 68,310  \\
1,338 & AS32491 Tucows.com          & 1 & 22,931  \\
1,284 & AS16844 Entrata             & 1 & 96,564 \\
946   & AS39570 Loopia AB           & 6 & 19,238  \\
824   & AS54635 Hillenbrand         & 1 & 117,251 \\
705   & AS53831 Squarespace         & 23& 386     \\
520   & AS12008 NeuStar             & 2 & 464     \\
516   & AS10668 Lee Enterprises     & 4 & 3,211   \\
\bottomrule

\end{tabular}
\label{table:high_privacy_hosting}
\end{table}

\begin{table}[t]
\centering
\caption{Top hosting providers with highest number of observed IP addresses.}
\begin{tabular}{rlrr}
\toprule
\small{Median} & \small{Organization}   & \small{Unique} & \small{Highest}\\
\small{\emph{k}} & \small{}   & \small{IPs} & \small{Rank}\\
\midrule
16 & AS13335 Cloudflare, Inc.       & 64,285 & 112  \\
 5 & AS16509 Amazon.com, Inc.       & 47,786 & 37   \\
 5 & AS46606 Unified Layer          & 27,524 & 1,265\\
 3 & AS16276 OVH SAS                & 22,598 & 621  \\
 3 & AS24940 Hetzner Online GmbH    & 21,361 & 61   \\
 4 & AS26496 GoDaddy.com, LLC       & 16,415 & 90   \\
 2 & AS14061 DigitalOcean, LLC      & 11,701 & 685  \\
 3 & AS14618 Amazon.com, Inc.       & 11,008 & 11   \\
 6 & AS32475 SingleHop LLC          & 10,771 & 174  \\
 2 & AS26347 New Dream Network      & 10,657 & 1,419\\
 7 & AS15169 Google LLC             &  9,048 & 1    \\
 3 & AS63949 Linode, LLC            &  8,062 & 2,175\\
 4 & AS8560  1\&1 Internet SE       &  6,898 & 2,580\\
 3 & AS32244 Liquid Web, L.L.C      &  6,412 & 1,681\\
 3 & AS19551 Incapsula Inc          &  6,338 & 1,072\\
 4 & AS36351 SoftLayer Technologies &  6,005 & 483  \\
 3 & AS16625 Akamai Technologies    &  5,862 & 13   \\
 4 & AS34788 Neue Medien Muennich   &  5,679 & 7,526\\
 6 & AS9371  SAKURA Internet Inc.   &  5,647 & 1,550\\
 3 & AS8075  Microsoft Corporation  &  5,360 & 20   \\
\bottomrule
\end{tabular}
\label{table:major_hosting}
\end{table}

In this section, we focus on the 81\% of multi-hosted domains that can benefit
from ESNI, and attempt to assess their actual privacy gain. Recall that a
website can gain some privacy benefit only if it is co-hosted with other
websites, in which case an on-path adversary will not know which among all
co-hosted websites is actually being visited. We use \emph{k}-anonymity to
model and quantify the privacy gain of multi-hosted domains.

Going back to Figure~\ref{fig:colocated_types}, we can apply this definition
in two ways, depending on whether we focus on IP addresses or domains. For a
given IP address, its \emph{k}-anonymity value (``\emph{k}'' for brevity)
corresponds to the number of co-hosted domains. For a given multi-hosted
domain, its \emph{k} may be different across the individual IP addresses on
which it is hosted, as the number of co-hosted domains on each of those
addresses may be different. Consequently, the \emph{k} value of a multi-hosted
domain is calculated as the median \emph{k} of all its IP
addresses.\footnote{Since most domains have similar \emph{k} values across
their hosting IP addresses, both mean and median can be used in this
case.} In both cases, the privacy gain increases linearly with \emph{k}. Based
on these definitions, we now explore the privacy gain of domains hosted on
different hosting and CDN providers.

Table~\ref{table:high_privacy_hosting} shows the top-ten hosting providers
offering the highest median \emph{k}-anonymity per IP address (i.e., greater
than 500). As shown in the third column, the average number of unique IP
addresses observed daily for each provider is very low, with half of them
hosting all domains under a single IP address.  Using the Hurricane
Electric BGP Toolkit, we confirmed that these providers are indeed small,
%hosting providers,
with many of them managing less than 10K IP addresses
allocated by regional Internet registries. When looking into the
popularity of the websites hosted by these providers, as shown in the last
column, the highest ranked website is only at the 386th position, hosted on
Squarespace, while more than half of these providers host websites that are
well below the top 10K.

Next, we investigate the \emph{k}-anonymity offered by major providers that
dominate the unique IP addresses observed. Table~\ref{table:major_hosting}
lists the top-20 major hosting and CDN providers with more than 5K unique
IP addresses observed. Unlike small hosting providers, these major providers
are home to more popular sites. Indeed, the most popular sites hosted by
these providers are all within the top 10K. In contrast to small providers,
however, the median \emph{k}-anonymity per IP address offered by these
providers is quite low, meaning that sites hosted on them will gain a much
lower level of privacy. Except from Cloudflare, which has the highest \emph{k}
of 16, all other providers have a single-digit \emph{k}.

Tables~\ref{table:high_privacy_hosting} and \ref{table:major_hosting}
represent two ends of the privacy spectrum for multi-hosted domains. On one
end, numerous but less popular domains are hosted on providers managing a
handful of IP addresses, benefiting from high \emph{k}-anonymity; on the other
end, fewer but more popular websites are hosted on providers managing a much
larger pool of millions of IP addresses, suffering from low
\emph{k}-anonymity. 

\begin{figure}[t]
\centering
\includegraphics[width=0.95\columnwidth]{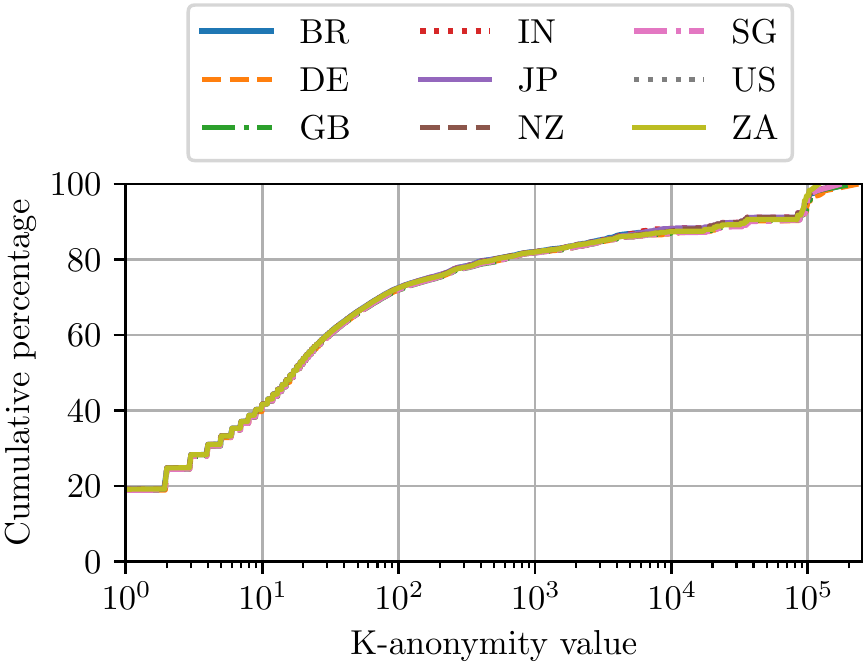}
\caption{CDF of the \emph{k}-anonymity for all studied domains across
  nine measurement locations (\emph{k}=1 corresponds to the 18.7\% of
  single-hosted domains).}
\label{fig:k_anonymity_cdf}
\end{figure}

To provide an overall view of the whole privacy spectrum,
Figure~\ref{fig:k_anonymity_cdf} shows CDFs of \emph{k} of all studied domains
across nine different regions. As illustrated, \emph{k} values are almost
identical across the nine regions from which we conducted our measurements.
While our DNS data shows that there are 471K (CDN-supported) domains served
from different IP addresses depending on the resolution location, the \emph{k}
values of these domains remain similar regardless of the DNS resolution
origin.

\begin{figure*}[t]
\centering
\includegraphics[width=1\textwidth]{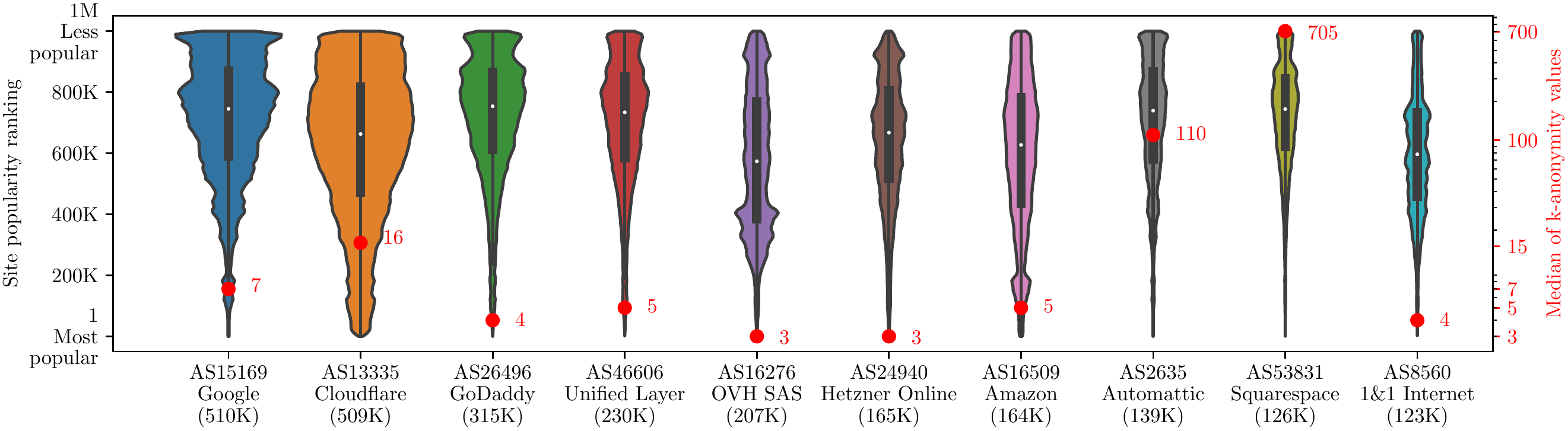}
\caption{Top providers that host most domains.}
\label{fig:populous_providers}
\end{figure*}

As discussed in \sectionref{sec:threat_model}, a low (e.g., single-digit)
\emph{k} cannot allow ESNI to offer meaningful privacy, given that i)~not all
\emph{k} sites will be equally popular, and ii)~website fingerprinting can be
used to improve attribution accuracy even further~\cite{touching12,
fingerprinting14, fingerprinting16, encrhttp06, Patil:2019}. Assuming that
\emph{k} needs to be greater than 100 to provide meaningful privacy, since an
adversary would correctly guess a domain being visited with a probability less
than 1\%, then according to Figure~\ref{fig:k_anonymity_cdf}, only about 30\%
of the sites will benefit from domain name encryption. We conduct a more
in-depth analysis of the probability with which an adversary would correctly
guess domains being visited based on Zipf's law
in~\sectionref{sub:quantify_weighted_privacy}.

Finally, we examine the top-10 providers that host the largest number of
domains among the ones studied. Although these mostly include some of the
providers listed in Table~\ref{table:major_hosting}, two of them are not
included on that table, and one (Squarespace) is actually included in
Table~\ref{table:high_privacy_hosting}. The violin plot of
Figure~\ref{fig:populous_providers} depicts the top-ten providers that host
most domains. The area of each violin is proportional to the number of domains
hosted by that provider, while the shape of each violin illustrates the
popularity ranking distribution of hosted websites. The median \emph{k} of
each provider is denoted by the red dot. Google and Cloudflare are the top
hosting providers, with more than 500K domains each. Other providers host
different numbers of domains, ranging from 315K to 123K.\footnote{Note that a
website may be hosted on more than one provider~\cite{Hoang2020:CCR}. In that
case, we count the site separately for each hosting provider.} Although
hosting fewer domains, both Automattic and Squarespace provide significantly
higher privacy with a \emph{k} of 110 and 705, respectively.

\subsection{Weighting the Privacy Benefit Based on Domain Popularity}
\label{sub:quantify_weighted_privacy}

In~\sectionref{sub:quantify_privacy}, we used the \emph{k}-anonymity model to
quantify the privacy benefit provided by multi-hosted domains. However, one
might consider that the model does not accurately capture a real-world
adversary, as not all co-hosted domains are equally popular. Adversaries could
base their guess on the probability that a domain is more (or less) likely to
be visited, according to the visit frequency of that domain compared to other
co-hosted domains. However, it is infeasible for us to obtain the data of
domain visit frequencies, since this is only known by the respective hosting
providers.

Fortunately, prior studies have shown that the relative visit frequency of
domains follows Zipf's law~\cite{zipf1929relative, Breslau1999INFOCOM}. More
specifically, Zipf's law states that the relative probability of a domain
($d$) being visited is inversely proportional to its popularity ranking ($P_d
\propto 1/(rank_d)^\alpha$). We thus apply Zipf's law\footnote{For simplicity,
we present results with $\alpha=1$, following the strict Zipf's law. However,
adjusting the value of $\alpha$ to match previous observations
\cite{Breslau1999INFOCOM} also gave similar results.} on the popularity
ranking of domains to compute the probability with which an adversary can
correctly guess that a  given domain is being visited.

\begin{figure}[t]
  \centering
  \includegraphics[width=0.95\columnwidth]{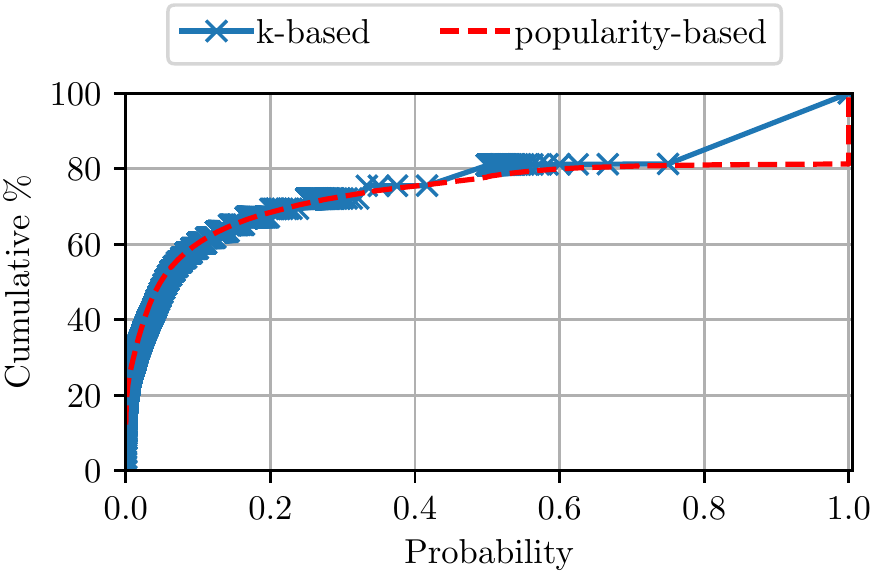}
  \caption{CDF of the probability of correctly guessing a visited domain
  based on the \emph{k}-anonymity value and popularity ranking information, as
  percentage of all tested domains.}
  \label{fig:probability_weighted_normal}
\end{figure}

From a privacy-detrimental IP address, it is straightforward for the adversary
to learn the domain being visited as the IP address solely hosts that single
domain. However, given a privacy-beneficial IP address that hosts multiple
domains, a more realistic adversary would make his guess based on the
probability that a domain is more likely to be visited compared to other
co-hosted domains. In order to compute this probability, we first obtain the
domains $d_1, \ldots, d_n$ that are hosted on a single $IP_j$ and compute
their $P_d$ values according to Zipf's law. We define $P_{d_{ij}} =
\frac{P_{d_i}}{\sum_{k=1}^{n} P_{d_k}}$ as the probability that domain $d_i$
was visited when $IP_j$ was observed.

For domains that are hosted on multiple IP addresses, the probability is
estimated by taking the median of all probabilities that the domain is visited
from all IP addresses hosting it. We therefore compute the probability that an
adversary can correctly guess domain $d_i$ that is hosted on $IP_1, \ldots,
IP_m$ as follows:
\begin{equation}
P_i = median(P_{d_{i1}}, \ldots, P_{d_{im}})
\end{equation}

As shown in Figure~\ref{fig:probability_weighted_normal}, our
\emph{k}-anonymity model is a close lower bound to the case where the
adversary considers the popularity rankings. The figure shows two CDFs of the
probability that the adversary can guess which domain is being visited. The
continuous (blue) line is computed based on the \emph{k}-anonymity value of
co-hosted domains. Each domain has an equal probability of $1/k$ to be
visited. The dashed (red) line is computed by applying the Zipf's law on the
domain popularity. We can see that even if adversaries rely on domain
popularity rankings to improve the accuracy of their prediction, the highest
probability that this guess is correct is similar to the probability estimated
by the \emph{k}-anonymity value.

\subsection{Domain-to-IP Mapping Stability}
\label{sec:ip_stability}

Besides the degree of co-hosting, the stability of a website's IP address(es)
also plays an important role in whether ESNI will provide meaningful privacy
benefits. If the IP address of a website changes quite frequently, this will
have a positive impact on the privacy offered due to ESNI. Unless adversaries
have enough resources to acquire all domain-to-IP mappings of interest at
almost real-time, they will no longer be able to use the destination IP
address as an accurate predictor of the visited website, because a previously
known domain-to-IP mapping may not be valid anymore. On the other hand,
mappings that remain stable over the time make it easier for adversaries to
monitor the visited websites.

In this section, we examine the stability of domain-to-IP mappings, and how it
affects privacy. We are particularly interested in finding how often
domain-to-IP mappings change. As discussed in~\sectionref{sec:method}, all top
lists of popular sites have their own churn (i.e., new sites appear and old
sites disappear from the lists on a daily basis). To prevent this churn from
affecting our analysis, we consider only the subset of domains that were
present daily on both top lists (\sectionref{sec:dataset}) during the whole
period of 61 days of our study. This set of domains comprises 2.6M domains,
from which we observed a total of 22.7M unique domain-to-IP mappings because a
domain may be hosted on hundreds of IP addresses.

\begin{figure}[t]
\centering
\includegraphics[width=0.95\columnwidth]{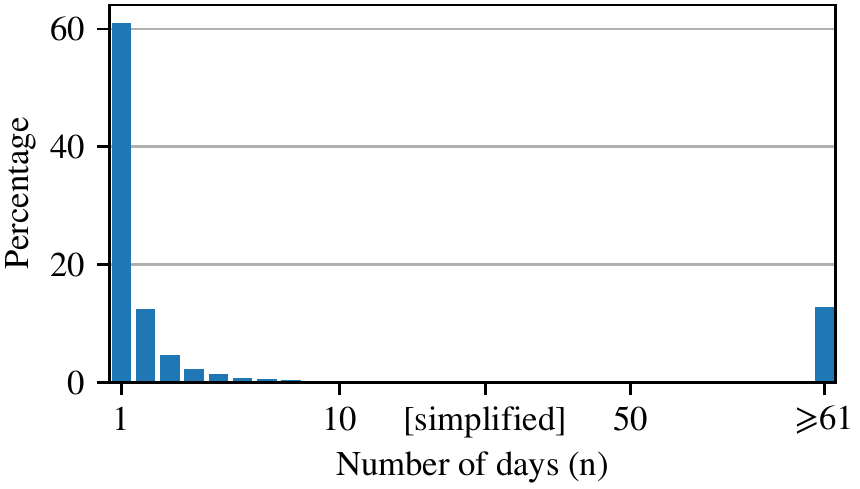}
\caption{Longevity distribution of domain-to-IP mappings as percentage of
number of mappings.}
\label{fig:domain_ip_longevity}
\end{figure}

Figure~\ref{fig:domain_ip_longevity} shows the distribution of the longevity
of these mappings in days. More than 80\% of the mappings last less than four
consecutive days (\emph{short-term mappings}), corresponding to 202K (7.7\%)
domains served from 400K unique IP addresses. On the other hand, 13\% of the
mappings remain unchanged for the whole study period (\emph{long-term
mappings}), corresponding to 2.4M domains served from 1.1M unique IP
addresses. As also shown in Figure~\ref{fig:domain_ip_longevity}, there are
two dominant clusters of domains that either change their hosting IP addresses
frequently or do not change at all. This is a favorable result for
adversaries, as it implies that they do not have to keep resolving a large
number of domains, since most domain-to-IP mappings remain quite stable over a
long period.

% \begin{figure}
%     \centering
    
%     \begin{subfigure}[b]{\columnwidth}
%         \includegraphics[width=0.9\columnwidth]{figures/domain_ip_longevity.pdf}
%         \caption{CDF of the longevity of domain-to-IP mappings.}
%         \label{fig:domain_ip_longevity_cdf}
%     \end{subfigure}

%     \begin{subfigure}[b]{\columnwidth}
%         \includegraphics[width=0.9\columnwidth]{figures/domain_ip_longevity_bar.pdf}
%         \caption{Longevity distribution of domain-to-IP mappings as percentage of number of mappings.}
%         \label{fig:domain_ip_longevity_bar}
%     \end{subfigure}

%     \caption{Longevity of domain-to-IP mappings.}
%     \label{fig:domain_ip_longevity}

% \end{figure}

\begin{figure}[t]
\centering
\includegraphics[width=0.95\columnwidth]{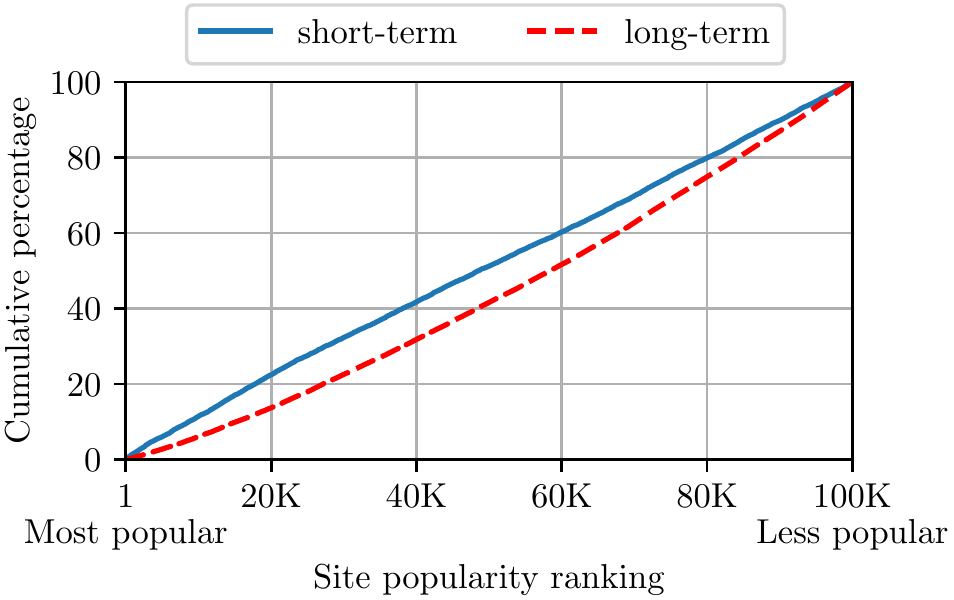}
\caption{CDF of domain popularity for short-term and long-term domain-to-IP
mappings.}
\label{fig:domain_rankings_distribution_of_mapping_longevity}
\end{figure}

The popularity distribution of the domains that correspond to these two
short-term and long-term mappings is shown in
Figure~\ref{fig:domain_rankings_distribution_of_mapping_longevity}. While
domains with short-term mappings are evenly distributed across the popularity
spectrum, domains with long-term mappings slightly lean towards lower
popularity rankings. This result can be attributed to the fact that more
popular websites are more likely to rotate their IP addresses for
load-balancing reasons, while less popular sites are more likely to be served
from static IP addresses.

\begin{figure}
\centering
\includegraphics[width=1\columnwidth]{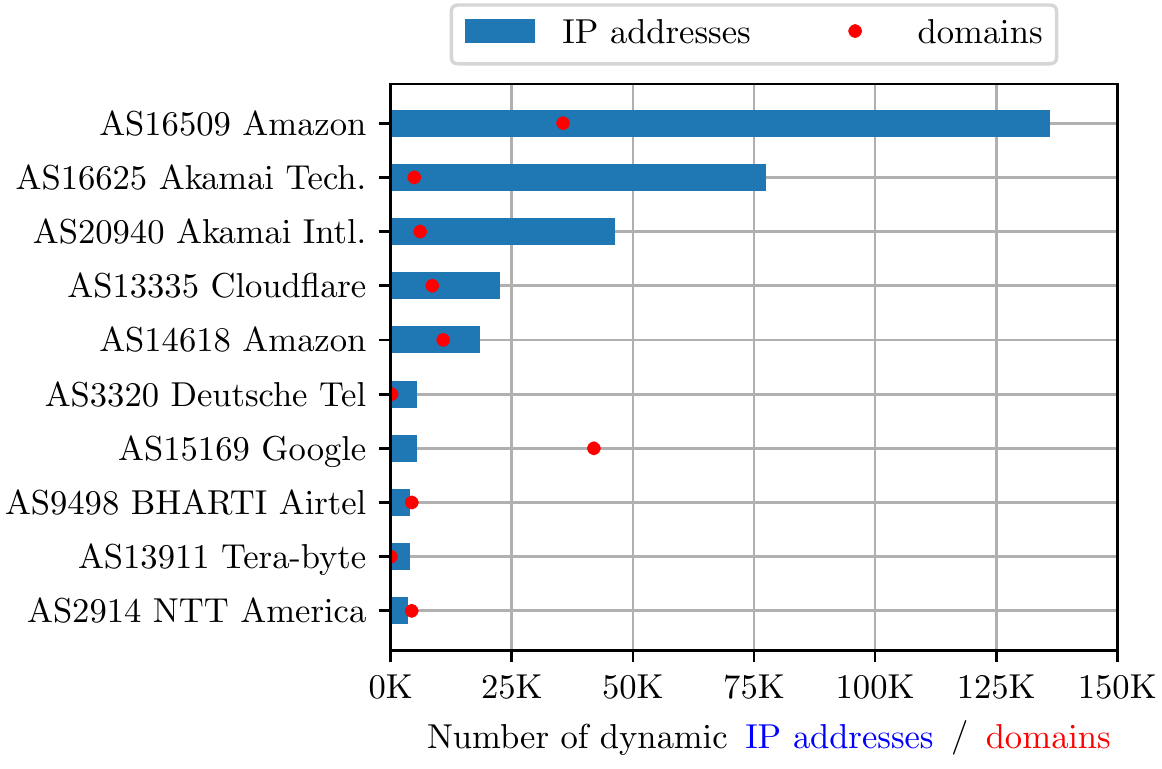}
\caption{Top providers with the highest number of high-churn IP addresses.}
\label{fig:10_most_dynamic_providers}
\end{figure}

An increased churn of IP addresses also helps ESNI provide better privacy. We
thus investigated which providers exhibit the highest churn rate by grouping
the IP addresses of short-term mappings according to their ASN.
Figure~\ref{fig:10_most_dynamic_providers} shows the top-ten providers with
the highest number of IP addresses in short-term mappings (bars). The dots
indicate the number of domains hosted on those IP addresses. Although Amazon
and Akamai do not top the list of providers that host most domains
(Figure~\ref{fig:populous_providers}), along with Cloudflare they occupy the
top five positions of the providers with the highest number of dynamic IPs.
Google uses a relatively small pool of around 5.3K IP addresses, to host more
domains (41K) than the other providers.

\section{Comparison with Other Datasets}
\label{sec:comparison}

In this section, we analyze existing public DNS datasets to examine the impact
of~i) larger datasets (in terms of number of domains), and~ii) more localized
vantage points, on the estimation of per-domain \emph{k}-anonymity. 

% In addition, we also analyze reverse DNS data to examine the current adoption
% status of reverse mapping from IP to domain name, and compare it to our number
% of single-hosted domains found in~\sectionref{sec:cohosted_analysis}, as this
% is another potential way for an adversary to infer a visited domain.

The Active DNS Project~\cite{Kountouras2016} is currently collecting A records
of about 300M domains derived from 1.3K zone files on a daily basis. In
addition to this effort, Rapid7~\cite{rapid7} also conducts active DNS
measurements at a large scale and offers researchers access to its data.
Unlike the Active DNS Project, Rapid7 resolves a much larger number of domains
(1.8B), but with a lower frequency (domains are resolved only once a week).
The dataset includes not only second-level domains from several TLD zone
files, but also lower-level domains obtained through web crawling and targeted
scanning with Zmap~\cite{Durumeric:2013}.

Different from these datasets, as discussed in \sectionref{sec:dataset}, our
domain name dataset is curated from the global lists of Alexa and Majestic. We
also perform measurements from vantage points around the world to observe localized DNS
responses from CDNs. In contrast, the above datasets are collected from local
vantage points, as their goal is to maximize the number of observed domains,
and not to exhaustively resolve all potential IP addresses of a domain. In
particular, the Active DNS Project is run from Georgia Tech, while Rapid7's
data is collected using AWS EC2 nodes in the US. To that end, we used two
datasets from the Active DNS Project and Rapid7 collected on March 29th, 2019
for our comparison. We sanitized poisoned records from these datasets, as
described in Appendix~\sectionref{sec:data_sanitization}, before analyzing
them.

%Understandably, their design choice, perhaps, is in favor of the number of
%domains resolved. It would be infeasible for them to conduct measurements
%from many vantage points at a global scale, which may result in high overhead
%costs to maintain dedicated infrastructures.

% \subsection{K-anonymity Comparison}
% \label{sec:data_comparison}

\begin{figure}[t]
\centering
\includegraphics[width=0.95\columnwidth]{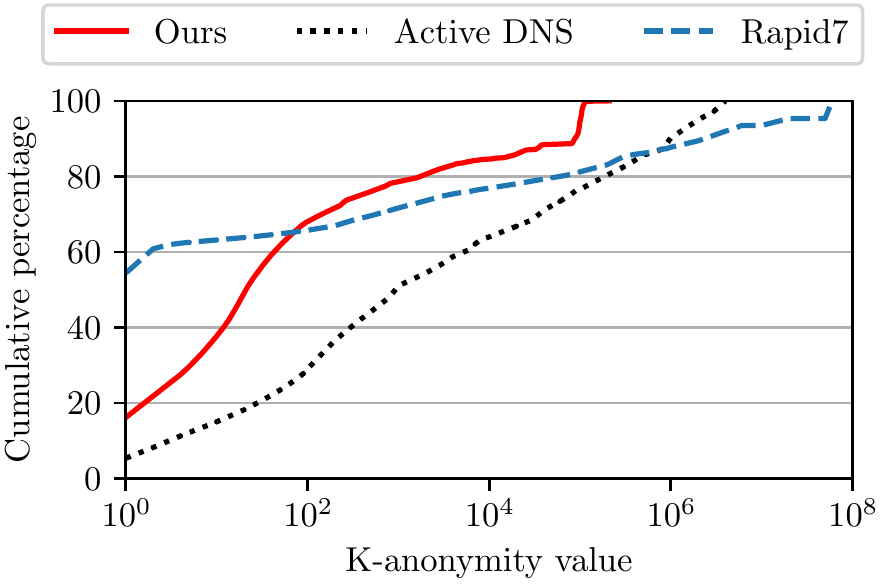}
\caption{CDF of the \emph{k}-anonymity value per domain as a percentage of
\textit{all domains} observed from the Active DNS Project, Rapid7, and our
datasets.}
\label{fig:others_k_anonymity}
\end{figure}

\begin{figure}[t]
\centering
\includegraphics[width=0.95\columnwidth]{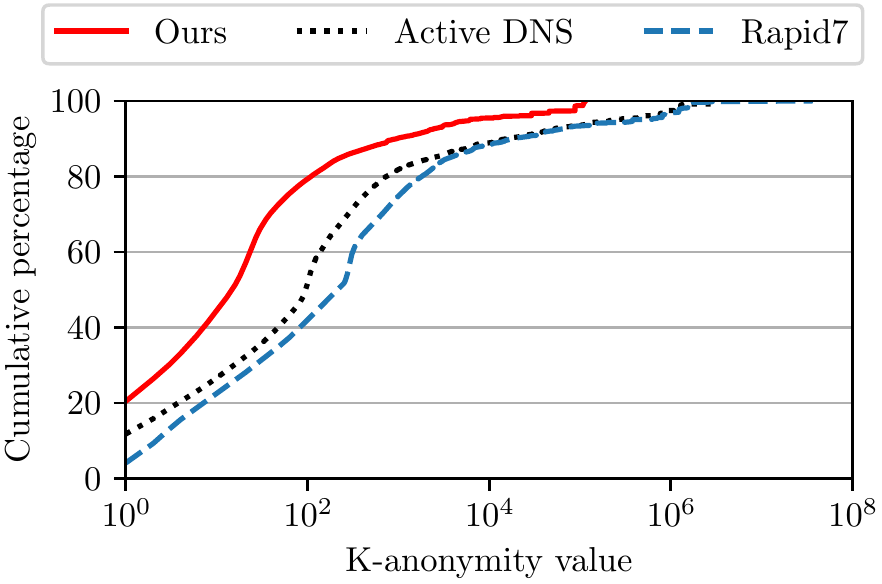}
\caption{CDF of the \emph{k}-anonymity value per domain as a percentage of
\textit{common domains} observed from the Active DNS Project, Rapid7, and our
datasets.}
\label{fig:common_domains_k_anonymity}
\end{figure}

Figure~\ref{fig:others_k_anonymity} shows the CDF of the \emph{k}-anonymity value per
domain for all the domains in the Active DNS, Rapid7, and our datasets,
while Figure~\ref{fig:common_domains_k_anonymity} shows the CDF of
the \emph{k}-anonymity value per domain for only common domains among the three
datasets.

In Figure~\ref{fig:others_k_anonymity}, when $\emph{k}=1$, there is some
difference in the percentage of single-hosted domains between the Active DNS
Project (5.3\%), Rapid7 (54.3\%), and our observation (18.7\%). As expected,
the percentage of single-hosted domains for Rapid7 is the highest because this
dataset is the largest (1.8B FQDNs), and includes lower-level FQDNs that may
host other services (e.g., email, DNS, SSH) instead of web content. On the
other hand, only 5.3\% of the domains in the Active DNS dataset are
single-hosted, since the dataset contains mostly A records of domains
extracted from TLD zone files, instead of many lower-level FQDNs.

When all domains are considered, our observation of single-hosted domains is
in between the two (18.7\%) because (as mentioned in~\sectionref{sec:dataset})
we derive our domain name dataset from the two global top websites lists
(Alexa and Majestic), and include not only second-level domains but also
lower-level FQDNs, as long as they are included in the lists and serve web
content. This aligns well with the results of Rapid7 and Active DNS, given the
inherent over-approximation of the Rapid7 dataset (due to non-web services and
highly unpopular or even single-use domains) and the inherent
under-approximation of the Active DNS Project dataset (due to only domains
extracted from TLD zone files).

When considering only common domains, the percentages of single-hosted domains
are 4.1\%, 11.7\%, and 20.4\% for the Rapid7, Active DNS, and our
dataset, respectively. In other words, some domains classified as
single-hosted in our dataset are actually multi-hosted when considering the
larger datasets. This result confirms our hypothesis discussed
in~\sectionref{sec:dataset} that by only considering the two sets of popular
websites, we would have missed those less popular, random, or even dormant
domains in the long tail. Thus, the result provided by our dataset can be
considered as a lower bound value of the actual \emph{k}-anonymity, since any
increase in \emph{k} provided by less well-known, random, or even dormant
domains is less meaningful from the perspective of adversaries whose goal is
to reveal a visited website by incorporating the popularity ranking
information in their ``guess.''

At the right end of the CDFs in both Figures~\ref{fig:others_k_anonymity}
and~\ref{fig:common_domains_k_anonymity}, the two larger datasets exhibit
\emph{k}-anonymity values higher than ours. The primary reason for this is
that these datasets include not only second-level domains, but also
third-level and longer FQDNs. The higher \emph{k} values also comprise the
long-tail of domains that are not included in our dataset, including less
popular domains, random or single-use FQDNs used for tracking, and malicious
domains~\cite{Szurdi:usenixsecurity14}.

\section{Discussion}
\label{sec:discussion}

\subsection{Recommendations}

While the \emph{security} benefits of DoH/DoT against on-path adversaries are
clear (e.g., prevention of MiTM or MoTS DNS poisoning attacks), our findings
show that Encrypted SNI alone cannot fully address the \emph{privacy} concerns
it aims to tackle. More effort and collaboration from all involved parties
(i.e., operators of DNS authoritative name servers, website owners, and
hosting and CDN providers) are needed. In this section we provide some
suggestions for maximizing the privacy benefits of ESNI.

\myparab{Full Domain Name Confidentiality.} In the current designs, plaintext
domain names are exposed through two channels: the SNI extension in TLS, and
traditional DNS name resolutions---the deployment of DoH/DoT is thus a
prerequisite for ESNI.
%Therefore, ESNI has to be used in combination with DoH/DoT to safeguard the
%domain name.
Equivalently, the use of DoH/DoT will not provide any meaningful
privacy if domain names are still exposed through the (unencrypted)
SNI extension in TLS handshake traffic.

Recently, there is a push for the deployment of DoH/DoT, with major
organizations already supporting it (e.g., Google, Cloudflare, Firefox),
though this has not been followed by an equivalent effort for the deployment
of ESNI. An even more complicated method of securing DNS traffic is
DNS-over-HTTPS-over-Tor, which has been already implemented and supported by
Cloudflare~\cite{cloudflareDNS-Tor}.
%On the contrary, while ESNI has been proposed together with TLS 1.3, it has
%not been widely adopted yet.
Unless the confidentiality of domain names is preserved on both channels (TLS
and DNS), neither technology can provide any actual privacy benefit if
deployed individually.

\myparab{Domain Owners.} Website owners who want to provide increased privacy
to their users can seek hosting providers or CDNs that offer an increased
ratio of co-hosted domains per IP address and/or highly dynamic domain-to-IP
mappings. In practice, however, this may be challenging. Our results show that
unfortunately only a few providers offer a high domain-to-IP ratio, while
other more pressing factors (e.g., site popularity) may tilt the decision
towards other more important factors, such as latency, bandwidth, or points of
presence.

While pointer (PTR) records are often not configured, from a privacy
perspective, their operation conflicts with DoH/DoT and ESNI as further
discussed in Appendix~\ref{sec:ptr_comparison}. Consequently, website operators should
not configure PTR records unless absolutely necessary (e.g., for email
servers). In addition, providers with a higher rotation of IP addresses are
more preferable, as this also helps in improving privacy.

\myparab{Hosting Providers.} Hosting and CDN providers are in a more
privileged position to achieve meaningful impact in helping improve the
potential privacy benefits of ESNI, as they can control the number of
co-hosted domains per IP address, and the frequency of IP addresses rotation.
Unless website owners prefer otherwise, providers could group more websites
under the same IP address (which, understandably, may not be desirable for
some websites).

To improve \emph{k}-anonymity even more meaningfully, providers should cluster
websites according to similarities in terms of traffic patterns and popularity
ranking, to hinder website fingerprinting attempts. As discussed
in~\sectionref{sec:ip_stability},  more dynamic hosting IP addresses can also
help improve visitors' privacy. Currently, the number of websites benefiting
from more short-lived domain-to-IP mappings is relatively small. While more
frequent IP address changes may complicate operational issues and are
certainly more challenging to deploy from a technical perspective (especially
for smaller providers), existing load balancing schemes already provide such a
capability, which could be tuned to also maximize privacy. In the future, it
may be worthwhile to explore more sophisticated schemes that actively attempt
to maximize privacy by increasing the ``shuffling'' rate of co-hosted domains
per IP address, to hinder attribution even further, especially when
considering more determined adversaries.

\subsection{Impact}

The deployment of domain encryption has various advantages and disadvantages
from the two---rather conflicting---perspectives of Internet censorship and
network visibility.

The existing plaintext exposure of domain names on the wire, as part of DNS
requests and TLS handshakes, has enabled the wide use of network traffic
filtering and censorship based on domain names~\cite{holdonDNS,
farnan2016cn.poisoning, china:2014:dns:anonymous, Pearce:2017:Iris,
Will:2016:Satellite, hoang:2019:measuringI2P}. In a future with all domain
name information being encrypted, DNS and SNI traffic will no longer be an
effective vector to conduct censorship. It is likely that censors will shift
to use IP-based blocking, which can be very effective if hosting IP addresses
of censored websites are stable and host only a handful of sites or
services~\cite{hoang:2018, Arun:foci18}.
% Place holder -- pointer to Appendix. In fact, we analyzed the Citizenlab
% global list and found that 90% of sensitive sites has one-to-one mapping.
However, if providers start adapting according to the above mentioned
recommendations, the cost of conducting IP-based blocking will increase, since
a censor will have to keep track of which IP address belong to which websites.

More importantly, the collateral damage caused by this type of blocking will
also increase dramatically if censored websites are co-hosted with multiple
other innocuous websites~\cite{Hoang2020:CCR}. Although some previous actions
from the side of providers (e.g., hindering domain
fronting~\cite{fifield2015blocking}) have shown that privacy is often given a
secondary priority~\cite{Fahmida2018}, as the collateral damage caused to
censors may also impact significantly the providers, the renewed recent focus
on privacy as a potential competitive advantage by some providers may
encourage the deployment of hosting schemes that will improve the privacy
benefits of ESNI.

On the other hand, while providing many security and privacy benefits, domain
name encryption can be a ``double-edged sword'' for network administrators who
want to have full visibility and control over domain resolutions in the
networks under their responsibility. Until now, the operation of firewalls,
intrusion detection systems, and anti-spam or anti-phishing filters has
benefited immensely from the domain name information extracted from network
traffic, as is evident by the series of works mentioned
in~\sectionref{sec:related_work} that employ DNS data to detect domain name
abuses and malicious online activities.

Under a full DoH/DoT and ESNI deployment, this visibility will be lost, and
systems based on domain reputation~\cite{Antonakakis:usenixsecurity10} and
similar technologies will be severely impacted. While many malicious domains
often hide themselves by sharing hosting addresses with other innocuous and
unpopular websites~\cite{Szurdi:usenixsecurity14}, it will be challenging to
detect and block them. A possible solution would be to rely solely on TLS
proxying using custom provisioned certificates, in order to gain back the
visibility lost by ESNI and DoH/DoT, which is already a common practice used
by transparent SSL/TLS proxies. Although this will defeat any privacy benefits
of these technologies, this may be an acceptable trade off for corporate
networks and other similar environments.

\section{Related Work}
\label{sec:related_work}

The domain name system is one of the core elements of the Internet and plays
an essential role for most online services. As a result, it has been (ab)used
for many different purposes. In this section, we review prior works that
investigate DNS from security and privacy perspectives, and some recent
studies that analyze the domain name ecosystem via empirical measurements.

From a security perspective, domain names have been heavily abused for illicit
purposes. For instance, domain squatting is one of the most common abuses. It
is used to register domains that are similar to those owned by well-known
Internet companies. Domain squatting has many variations, including
typo-squatting~\cite{Agten:NDSS17, Szurdi:usenixsecurity14, Khan:SP15},
homograph-based squatting~\cite{Gabrilovich:2002, Quinkert:CNS19},
homophone-based squatting~\cite{Nikiforakis:2014},
bit-squatting~\cite{Nikiforakis:WWW13}, and
combo-squatting~\cite{Kintis:CCS17}. Domain names registered using these
squatting techniques can then be used for phishing~\cite{Pariwono:ASIACCS18,
Quinkert:CNS19} or distributing malware~\cite{Alowaisheq:NDSS19}. To cope with
these unwanted domain names, DNS data has been used intensively to create
domain name reputation systems to detect
abuse~\cite{Antonakakis:usenixsecurity10, Antonakakis:usenixsecurity11,
Antonakakis:usenixsecurity12, Krishnan:DSN11}.

Another major form of DNS abuse is DNS poisoning, in which an on-path observer
can easily observe and tamper with DNS responses to redirect users to
malicious websites or to censor unwanted content~\cite{lowe2007great,
holdonDNS, farnan2016cn.poisoning, china:2014:dns:anonymous, Pearce:2017:Iris,
Will:2016:Satellite}. The exposure of domain names in DNS requests and TLS
handshakes (due to SNI) has also been extensively used for traffic filtering
and censorship~\cite{zchai2019foci, hoang:2019:measuringI2P}.

As mentioned in~\sectionref{sec:dns}, the traditional design of DNS exposes
Internet users to severe privacy risks. In addition to on-path observers
(discussed in~\sectionref{sec:threat_model}), previous works have also studied
the privacy risk associated with centralizing all domain name resolutions to
third-party recursive resolvers (e.g., 8.8.8.8, 1.1.1.1)~\cite{Zhao:2007:PIR,
Castillo2008, Castillo2009, Shulman:2014, Herrmann2014:EncDNS,
Kintis:DIMVA16}. Zhao et al.~\cite{Zhao:2007:PIR} propose to add random noise
and use private information retrieval to improve privacy by obfuscating DNS
queries. These proposals however have turned out to be impractical and
insecure under certain circumstances~\cite{Castillo2009}, and have not  been
adopted. Lu et al.~\cite{Lu2010} propose privacy-preserving DNS (PPDNS), which
is based on distributed hash tables and computational private information
retrieval. More recently, Hoang et al.~\cite{Hoang2020:MADWeb} propose
\emph{K}-resolver as a mechanism to distribute DoH queries among multiple
recursors, thus exposing to each recursor only a part of a user's browsing
history. Sharing similar goals with our study, Shulman et
al.~\cite{Shulman:2014} examined the pitfalls of DNS encryption. By analyzing
the co-residence of zone files on name servers, the authors argue that
guessing visited domains by destination IP address does not provide a
significant advantage. Our findings, however, show that this is only the case
for a small number of domains that are co-hosted with an adequate number of
other domains.

% There have been numerous efforts in encrypting DNS queries between a client
% and a DNS server~\cite{dempsky2010dnscurve, denis2015dnscrypt,
% bernstein2009dnscurve, wijngaards2014confidential}. Moreover, the DNS over
% HTTPS (DoH)~\cite{rfc8484} and DNS over TLS (DoT)~\cite{rfc7858} RFCs were
% also proposed to address the privacy issues of the DNS protocol. 

Hounsel et al.~\cite{hounsel2019analyzing} study the effect of DoH/DoT on
performance of domain name resolution and content delivery. The study finds
that the resolution time of DoH/DoT is longer than traditional DNS resolution.
Of the two new technologies, DoT provides better page load times while DoH at
best has the same page load times as DNS. They also find that DoT and DoH
perform worse than DNS in networks with sub-optimal performance. Similarly, a
recent study by Bottger et al.~\cite{bottger2019imc} analyzes the DoH
ecosystem and shows that they can obtain more advanced privacy features of DoH
with marginal performance degradation in terms of page load times.

% It is important to note that a client using DoH/DoT along with TLS 1.2 will
% still leak the domain name being visited from the SNI field. Thus, the RFC
% for Encrypted SNI (ESNI)~\cite{rfc-draft-ietf-tls-esni-03} has been proposed
% to be used as an extension to TLS 1.3 that will encrpyt the SNI field. From
% a censorship perspective, Frolov et al.~\cite{frolov2019use} compare
% real-world TLS traffic and the TLS implementations of popular censorship
% circumvention tools and find that the traffic from these TLS configurations
% are distinguishable from real traffic. Therefore, censors can identify
% clients based on the clear text Client Hello messages. A recent study on I2P
% censorship~\cite{hoang:2019:measuringI2P} indicates that the most dominant
% I2P filtering technique is based on domain names which are visible to the
% censor in DNS queries and the SNI extension. Chai et
% al.~\cite{zchai2019foci} also characterizes SNI-based censorship in China
% and measures the prevalence of ESNI for the Alexa top million websites.
% Their results indicate that ESNI is in its early states and only 10.9\% of
% the Alexa top million websites support it.}

There have also been studies that investigated the robustness of the DNS
ecosystem through various types of measurements. Ramasubramanian et
al.~\cite{Ramasubramanian:imc05} leverage a dataset of almost 600K domains to
study their trusted computing base, which is the set of name servers on which
a FQDN is hosted. The study shows that a typical FQDN depends on 46 servers on
average. Dell'Amico et al.~\cite{Dell'Amico:ACSAC17} use DNS data collected
through both active and passive measurements to also investigate the ecosystem of
dependencies between websites and other Internet services. Similarly to our
work, Shue et al.~\cite{Shue:2007} use DNS data collected by both passive and
active measurements to study web server co-location and shared DNS
infrastructure. However, their measurements were conducted from a single
location, while excluding all servers belonging to CDNs. Furthermore, the
passive DNS dataset used was collected by capturing network traffic from the
authors' institute, therefore facing all potential issues of a passive
measurement discussed in~\sectionref{sec:dns_measurement}. More recently,
Hoang et al.~\cite{Hoang2020:CCR} revisit the results
of Shue et al.~\cite{Shue:2007} by conducting a
large-scale active DNS measurement study, which reveals that the Web is still
centralized to a handful of hosting providers, while IP blocklists cause less
collateral damage than previously observed regardless of a high level of
website co-location.

\section{Conclusion}
\label{sec:conclusion}

The deployment of encrypted SNI in TLS, combined with DNS over HTTPS/TLS,
will definitely provide many security benefits to Internet users.
However, as we have shown in this work, a significant effort is still
needed in order for these same technologies to provide meaningful privacy
benefits. More specifically, while domain name information is encrypted, the
IP address information is still visible to any on-path observers and can be
used to infer the websites being visited.

Using DNS data collected through active DNS measurements, we
studied the degree of co-hosting of the current web, and its implications in
relation to ESNI's privacy benefits. Quantifying these benefits for co-hosted
websites using \emph{k}-anonymity,
we observed that the majority of popular websites (about half of all domains
studied) will gain only a small privacy benefit (\emph{k}<16). Such a small
degree of co-hosting is not
enough to withstand determined adversaries that may attempt to perform
attribution by considering the popularity or even the traffic patterns of the
co-hosted websites on an observed destination IP address.
Domains that will obtain a more meaningful privacy benefit
(\emph{k}>500) include only vastly less popular websites mostly hosted by
smaller providers, while 20\% of the websites, will not gain any benefit at
all due to their one-to-one mapping between domain name and hosting IP address.

We hope that our findings will raise awareness about the remaining effort that
must be undertaken to ensure a meaningful privacy benefit from the deployment
of ESNI. In the meantime, privacy-conscious website owners may seek hosting
services offered by providers that exhibit a high ratio of co-hosted domains
per IP address, and highly dynamic domain-to-IP mappings.

\section*{Acknowledgements}

We are grateful to Manos Antonakakis, Panagiotis Kintis, and Logan O'Hara from
the Active DNS Project for providing us their DNS dataset, and to Rapid7 for
making their datasets available to the research community.

We would like to thank all the anonymous reviewers for their thorough feedback
on earlier drafts of this paper. We also thank Hyungjoon Koo, Shachee Mishra,
Tapti Palit, Seyedhamed Ghavamnia, Jarin Firose Moon, Christine Utz,
Shinyoung Cho, Rachee Singh, Thang Bui, and others who preferred to remain
anonymous for helpful comments and suggestions.

This research was supported in part by the National Science Foundation under
awards CNS-1740895 and CNS-1719386. The opinions in this paper are those of
the authors and do not necessarily reflect the opinions of the sponsors.

\bibliographystyle{ACM-Reference-Format}
\balance
\bibliography{bibliography}
\appendix

\begin{table}[t]
\centering
\caption{Breakdown of the five largest TLDs studied.}
\vspace{-0.8em}
\begin{tabular}{lrr}
\toprule
      & \textbf{Daily} & \textbf{Total} \\ [0.5ex]
\midrule
TLDs  &          1,031 &          1,125 \\
FQDNs &      7,556,066 &     13,597,409 \\
\hline
.com  &      3,835,080 &      7,026,005 \\
.org  &        347,993 &        584,924 \\
.de   &        264,057 &        501,597 \\
.net  &        263,262 &        442,729 \\
.ru   &        210,701 &        346,194 \\
\bottomrule
\end{tabular}
\label{tab:tld_breakdown}
\end{table}

\setlength{\textfloatsep}{0.3cm}

\section{Breakdown of domains studied}
\label{sec:tld_breakdown}

As shown in Table~\ref{tab:tld_breakdown}, our derived list comprises an
average of 7.5M popular fully qualified domain names (FQDNs) collected on a
daily basis, covering 1,031 TLDs. For the whole experiment duration, we
studied a total of 13.6M domains and 1,125 TLDs. Table~\ref{tab:tld_breakdown}
also shows the top five largest TLDs in our dataset, with \texttt{.com} being
the most dominant, comprising more than 50\% of the domains observed.

\section{Poisoned Response Sanitization}
\label{sec:data_sanitization}

While processing public DNS datasets from other sources (to which we compare
our findings in \sectionref{sec:comparison}), we surprisingly discovered
thousands of low-ranked or obscure domains seemingly being co-hosted on the
same IP addresses that also host very popular websites, such as Facebook and
Twitter---which of course was not actually the case. As part of our
investigation, we observed that the authoritative servers of most of these
domains were located in China (using the MaxMind dataset~\cite{maxmind}). We
then queried the same domains from outside China using their authoritative
servers, and indeed received responses pointing to IP addresses that belong to
either Facebook or Twitter. By inspecting network traffic captures taken
during these name resolutions, we observed that the initial response
containing the wrong (falsified) IP address was followed by another DNS
response with the same valid DNS query ID that contained a different (correct)
IP address.

We attribute the above observed behavior to DNS-based censorship by the
``Great Firewall'' (GFW) of China~\cite{Zittrain2003, lowe2007great,
claytonchina, china-gfw, anon2012cn.collateral, wright2014cn.regional}, which
has also been observed and analyzed by previous
studies~\cite{china:2014:dns:anonymous, farnan2016cn.poisoning}. Censorship
leakage happens due to the GFW's filtering design, which inspects and censors
both egress and ingress network traffic. While some censors (e.g., Pakistan,
Syria, Iran) forge DNS responses with NXDOMAIN~\cite{rfc8020, syriacensorship,
pakistancensorship, Chung:2016:IMC16} or private addresses~\cite{Aryan:2013},
making them easier to distinguish, China poisons DNS responses with routable
public IP addresses belonging to other non-Chinese
organizations~\cite{china-gfw, farnan2016cn.poisoning,
hoang:2019:measuringI2P}. In contrast to the findings of previous works,
however, in this case the real hosting IP addresses of the censored domains
are located within China, while previous works mostly focus on investigating
the blockage of websites that are hosted outside China (e.g., google.com,
facebook.com, blogger.com).

To validate our findings, we cross-checked the IP addresses from second (real)
DNS responses with the ones obtained by resolving the same domains from
locations in China. As the authoritative servers of these domains are also in
China, our queries did not cross the GFW, which mostly filters traffic at
border ASes~\cite{crandall2007.conceptdoppler, xu:2011:china,
china:2014:dns:anonymous}, and thus were not poisoned.

%Investigating further this irregular blocking behavior and understanding the
%motivation behind it can be part of future work on Internet censorship. For
%example, it would be interesting to investigate the motivation behind
%poisoning the DNS responses of sites hosted in China, where authorities
%already have the privilege to take them down. Nevertheless, this is beyond
%the scope of this paper, in which we only focus on obtaining an accurate
%dataset of co-hosted domains, and thus must avoid falsely attributing
%poisoned domains as co-hosted with other legitimate domains.

We follow this verification technique, where we issue additional queries to
resolve domains whose authoritative name server is located in China and then
detect injected DNS packets to exclude poisoned responses from analyses. In
total, we detected more than 21K domains based in China with poisoned
responses. Table~\ref{table:poinsoned_asn_ips} shows the top /24 IP subnets
belonging to Facebook, Twitter, and SoftLayer, which are the most frequently
observed in poisoned responses. Our observation aligns with recent findings of
other censorship measurement studies~\cite{iclab_SP20,
hoang:2019:measuringI2P}.

\begin{table}[t]
\centering
\caption{Most frequently abused subnets in poisoned DNS responses from China.}
\vspace{-0.8em}
\begin{tabular}{lll}
\toprule
\textbf{AS32934} & \textbf{AS13414}   & \textbf{AS36351} \\
\textbf{Facebook} & \textbf{Twitter}   & \textbf{SoftLayer} \\
\midrule
31.13.72.0                & 199.59.148.0               & 74.86.12.0                 \\
31.13.69.0                & 199.59.149.0               & 67.228.235.0               \\
31.13.73.0                & 199.16.156.0               & 74.86.151.0                \\
31.13.66.0                & 199.59.150.0               & 75.126.124.0               \\
69.171.245.0              & 199.16.158.0               & 67.228.74.0                \\
\bottomrule
\end{tabular}
\label{table:poinsoned_asn_ips}
\end{table}

%Sampling one of the poisoned domains, e.g., \emph{shenmakan.com}, we find
%that prior works' datasets contain several poisoned responses of the domain.
%We further discover that many commercial services, that provide historical
%DNS data, also observe poisoned records of this domain in their
%datasets~\cite{dns_poison_securitytrail, dns_poison_virustotal}. Similar with
%our observation, IP addresses of \emph{shenmakan.com} recorded by these
%services also belong to either \emph{Facebook}, \emph{Twitter}, or
%\emph{SoftLayer}. Hence, our finding means that any DNS measurement conducted
%from outside China to resolve these China-based domains could have been
%affected because all DNS queries would have to pass through the GFW to reach
%their authoritative name servers, thus being tampered. 

\section{Reverse DNS Lookups}
\label{sec:ptr_comparison}

The Internet Engineering Task Force (IETF) recommends that it should be
possible to conduct a reverse DNS lookup for every given
domain~\cite{rfc2317}. In a forward DNS lookup, a domain name is resolved to
an A (IPv4) record, while a reverse DNS lookup sends out an IP address to ask
for its associated FQDN. The DNS record storing this information is called a
PTR (pointer) record. Unless configured to point to a FQDN by its owner, it is
not compulsory to configure PTR records for every IP address.

%PTR records often store the IP address with its segments reversed, suffixed
%by \emph{.in-addr.arpa} at the end~\cite{rfc2317}. For example, if a domain
%has an IP address of 1.2.3.4, the PTR record stores the information as
%\emph{4.3.2.1.in-addr.arpa.}

Although performing reverse DNS lookups seems to be a straightforward way of
mapping a given IP address back to its associated FQDN, this can potentially
uncover only single-hosted domains. More importantly, not all reverse DNS
queries return a (meaningful\footnote{Many PTR records are formatted with
dash-separated IP segments. For example, the Amazon EC2 IP address
\emph{54.69.253.182} has a PTR record to
\emph{ec2-54-69-253-182.us-west-2.compute.amazonaws.com}, which may not
actually correspond to a user-facing web service.}) domain name because the
IETF's recommendation is only optional, and thus not adopted universally.

We analyzed Rapid7's reverse DNS dataset, which contains PTR records for the
whole public IPv4 space, to see how many IP addresses could be used to reveal
the visited destinations under the assumed global ESNI deployment, by just
performing a reverse DNS lookup. We find that there are 1.27B unique IP
addresses having PTR records. Of these, at least 172M (14\%) of them point to
a meaningful FQDN (i.e., not in the form of dash-separated IP segments).
Within this set of domains, we could find 136K single-hosted domains observed
by our dataset. This means about 10\% of single-hosted domains have PTR
records configured. Under a global ESNI deployment, IP addresses of these
domains would be detrimental to the privacy of users who connect to them.	

As expected, we also observed more than 3M PTR records in which domain names
explicitly indicate through the prefix ``mail'' that they correspond to email
servers. These email servers support PTR record because many providers will
not accept messages from other mail servers that do not support reverse
lookups~\cite{rfc2505}.

\end{document}